\documentclass{INTERSPEECH2023}

\interspeechcameraready

\title{Flowchase: a Mobile Application for Pronunciation Training}
\name{Noé Tits$^1$, Zoé Broisson$^1$}
\address{
  $^1$Flowchase, Belgium
  }
\email{noe@flowchase.app, zoe@flowchase.app}

\begin{document}

\maketitle
 
\begin{abstract}
In this paper, we present a solution for providing personalized and instant feedback to English learners through a mobile application, called Flowchase, that is connected to a speech technology able to segment and analyze speech segmental and supra-segmental features.
The speech processing pipeline receives linguistic information corresponding to an utterance to analyze along with a speech sample. After validation of the speech sample, a joint forced-alignment and phonetic recognition is performed thanks to a combination of machine learning models based on speech representation learning that provides necessary information for designing a feedback on a series of segmental and supra-segmental pronunciation aspects.

\end{abstract}

\section{Introduction}

In the field of Computer-Assisted Language Learning (CALL), there are nowadays still very few solutions focusing on oral skills, and specifically on pronunciation. Computer-Assisted Pronunciation Training (CAPT) is an important research discipline, but there is a lack of concrete applications, although explicit focus on pronunciation, when combined with the use of technologies, has a significant impact on L2 learners pronunciation~\cite{flowchase_study-22-cordier, ELSA-anguera2016english}. A reason for this situation is the gap of complexity between developing feedback on written, reading or listening skills compared to spoken skills. Indeed for the first three skill sets, implementing simple heuristics based on multiple answer exercises, or matching a user answer to a gold standard is straightforward. On the contrary, providing feedback on spoken skills is not. A speech technology tailored to analyzing segmental and supra-segmental patterns is necessary. 

The techniques of mispronunciation errors have been close to the findings of speech recognition area, from HMM-GMM~\cite{gop_gmm_hmm-00-witt}, to DNN-HMM~\cite{gop_dnn_hmm-15-hu} and more recently, transformers~\cite{gop_transformers-gong2022transformer}. Indeed the tasks share a strong common characteristic, which is extracting information from audio, a representation of human speech, be it text or phonetics. 

Transfer Learning~\cite{tan2018survey_deep_transfer_learning} is today a widely used technique in Deep Learning for leveraging models trained on related tasks for which there exist abundant datasets towards tasks for which few data exist.
This principle has been applied successfully for speech technology application~\cite{wang2015transfer} with few available data such as speech recognition for low resource languages, emotion recognition in speech~\cite{asr-based-features-18-tits}, emotional or expressive speech synthesis~\cite{exploring_transfer_learning-19-tits, visualization-19-tits} or voice conversion~\cite{zhou2022emotional}, and also to pronunciation assessment~\cite{lin2021deep}.
A specific form of Transfer Learning that was shown to be very efficient is self-supervised learning where a model is trained to learn representations of input data without the need for explicit supervision.

In this paper, we present a complete system able to provide a pronunciation training based on a speech technology built on top of a wav2vec2~\cite{baevski2020wav2vec} model adapted for mispronunciation detection, integrated in a mobile application. Although the application contains a mix of tutorials, listening activities and speaking activities, we focus here on the description of the speaking activities that involves the speech processing pipeline for analyzing English learners' pronunciation and providing feedback.




  



\section{System}

\begin{figure}[t]
  \centering
  \includegraphics[width=1\linewidth]{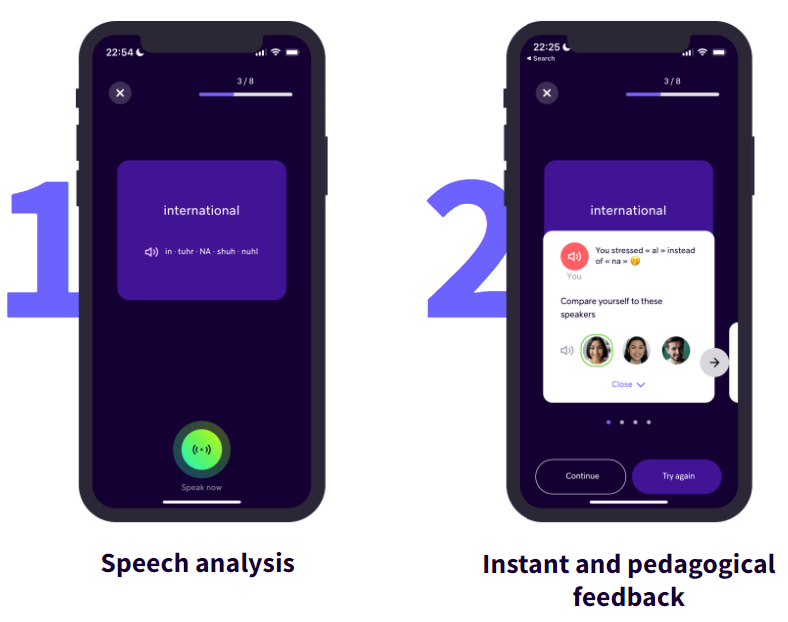}
  \captionsetup{skip=5pt,  belowskip=-15pt} 
  \caption{Sequence of steps in a Speaking Activity}
  \label{fig:app_steps}
\end{figure}

Figure~\ref{fig:app_steps} describes the main steps of the user experience inside a speaking exercise of a learning program. First, the exercise data is shown to the user. Specifically, it shows an English utterance that the user is expected to say, with a pronunciation guide to help him understand how it has to be pronounced. The pronunciation of the sentence can also be heard thanks to a set of different actor recordings with different English variations. 
On this screen, the user can record himself. Then the audio recording is sent to the speech technology backend along with the exercise information in order to perform segmentation and analysis of the speech sample. From this analysis, a number of information are extracted depending on the pronunciation aspect analyzed.

In the second screen, feedback cards are shown to the user in order to communicate the result of the analysis, and advice in order to improve.

\begin{figure}[t]
  \centering
  \includegraphics[width=1.1\linewidth]{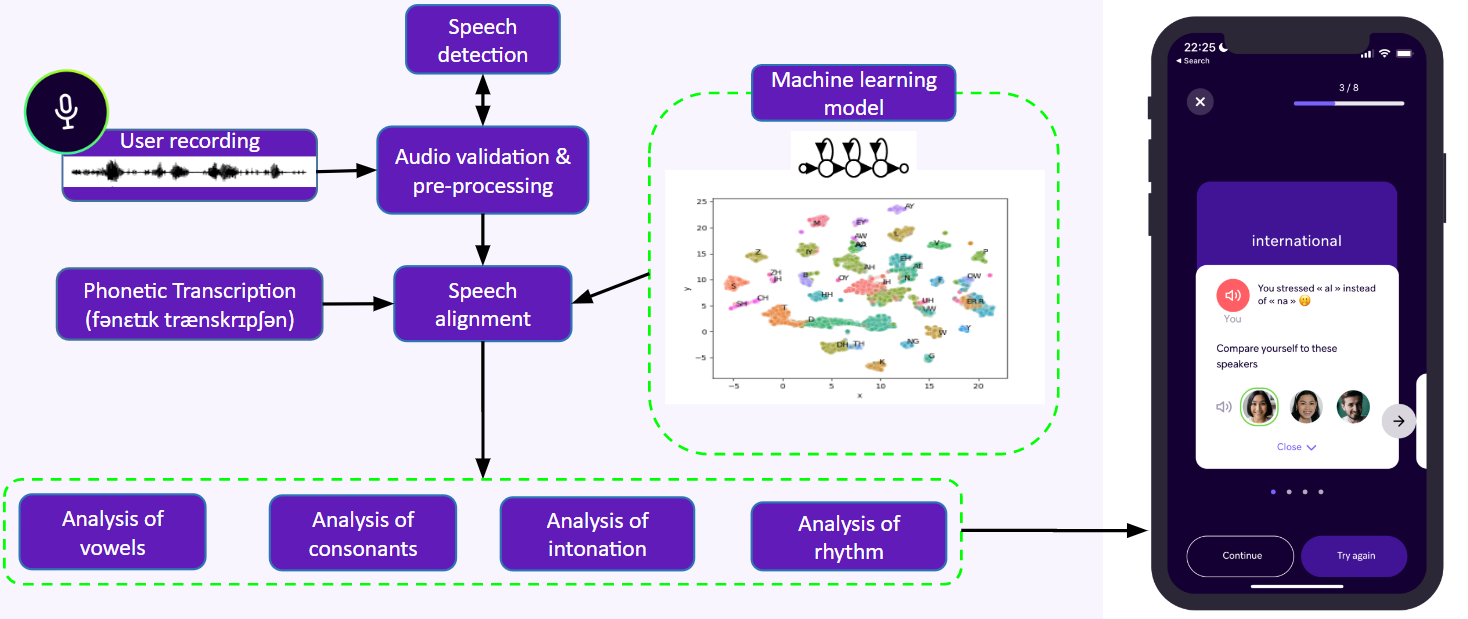}
  \captionsetup{skip=5pt,  belowskip=-5pt} 
  \caption{Description of the processing steps of a user recording towards pronunciation feedback}
  \label{fig:app_tech_diagram}
\end{figure}

Figure~\ref{fig:app_tech_diagram} details the processing steps happening in the second step explained above. The speech analysis takes as inputs exercise information such as the phonetic content and the English learner's speech sample. The user recording has first to be validated thanks to a series of test on audio that checks is a valid speech sample, including:

\begin{itemize}
    \item the duration of the audio is plausible to have a human speech rate compared to the expected utterance
    \item the speech sample contains voiced content
    \item the phonetic content in speech is sufficiently close from the phonetic content
\end{itemize}

\begin{figure}[t]
  \centering
  \includegraphics[width=\linewidth]{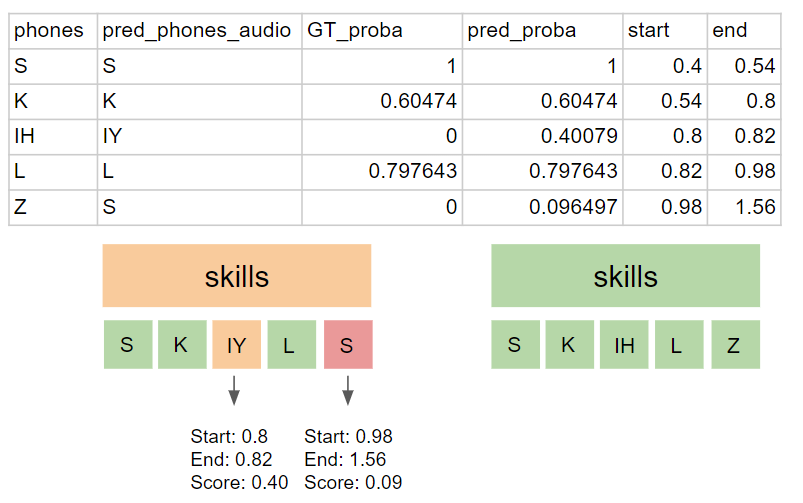}
    \captionsetup{skip=5pt,  belowskip=-15pt} 
  \caption{Pronunciation feedback on a word}
  \label{fig:app_feedback_skills}
\end{figure}

If the speech sample is validated, a combination of machine learning models based on speech representation learning is used for performing a forced-alignment between the speech sample and the phonetic transcription in order to extract the start and end timings of each phoneme of the sequence. The machine learning model also analyzes the phonetic content of the audio and allows us to extract information related to set of different pronunciation aspects such as analysis of vowels or consonants, and specifically analyzing minimal pairs, as shown in Figure~\ref{fig:app_feedback_skills}, analysis of intonation such as word stress or sentence stress, and other supra-segmental aspects like an analysis of pauses between breath groups in an utterance.

An example of analysis results on a word from a sentence is shown in Figure~\ref{fig:app_feedback_skills}. Expected phonemes and predicted phonemes are extracted along with the start and end timings, as well as the respective posterior probabilities according to the statistical model.

\section{Conclusions}

In this paper, we presented Flowchase, a mobile application for personalized pronunciation training that utilizes a speech technology pipeline for analyzing English learners' pronunciation and providing instant feedback. We employed transfer learning and self-supervised learning techniques to build a speech technology model for detecting mispronunciations based on the wav2vec2 architecture. 

The system provides feedback on both segmental and supra-segmental aspects of pronunciation. Our solution addresses the gap in current computer-assisted language learning applications, which mostly focus on written, reading, or listening skills. Flowchase provides a tool for improving oral language skills, particularly pronunciation, which is crucial for effective communication. Future work includes testing the effectiveness of the application and the speech technology pipeline in real-world settings and extending the system to support other languages.

\vspace{-5pt}

\section{Acknowledgements}
This work is part of the project \textit{REDCALL} that is partially funded by a FIRST Entreprise Docteur program from SPW Recherche\footnote{https://recherche.wallonie.be/}

\vspace{-10pt}

\bibliographystyle{IEEEtran}
\bibliography{mybib}

\end{document}